\documentclass[prd,twocolumn,notitlepage,showpacs,preprintnumbers,amsmath,amssymb,nofootinbib,APS,10pt,superscriptaddress]{revtex4-1}

\usepackage{dcolumn}
\usepackage{bm}
\usepackage[applemac]{inputenc}
\usepackage[spanish,english]{babel}
\usepackage{amsmath,amssymb,amsfonts,latexsym,cancel}
\usepackage{graphicx}
\usepackage{color}
\usepackage{soul}
\usepackage{ulem}
\usepackage{hyperref}

\allowdisplaybreaks

\newcommand{\be}{\begin{equation}}
\newcommand{\ee}{\end{equation}}
\newcommand{\bea}{\begin{eqnarray}}
\newcommand{\eea}{\end{eqnarray}}

\begin{document}
\title{{\bf Running gravitational  couplings, decoupling,  and curved spacetime renormalization }}

\author{Antonio Ferreiro}\email{antonio.ferreiro@ific.uv.es}

\author{Jose Navarro-Salas}\email{jnavarro@ific.uv.es}
\affiliation{Departamento de F\'isica Te\'orica and IFIC, Centro Mixto Universidad de Valencia-CSIC. Facultad de F\'isica, Universidad de Valencia, Burjassot-46100, Valencia, Spain.}

\begin{abstract}
We propose to slightly generalize the DeWitt-Schwinger adiabatic renormalization subtractions  in curved space to  include an arbitrary renormalization mass scale $\mu$. The new predicted running for the gravitational couplings are fully consistent with decoupling of heavy massive fields. 
This is a somewhat improvement with respect to the more standard treatment of minimal (DeWitt-Schwinger) subtractions via dimensional regularization.  
  We also show how the vacuum metamorphosis model emerges from the  running couplings.   \\

 \end{abstract}

\date{\today}
\maketitle

 \section{Introduction} 
One of the cornerstones in quantum field theory has been the design of regularization/renormalization schemes that allows us to overcome ultraviolet divergences when computing physical  observables \cite{qftbook1, qftbook2, qftbook3}. In perturbative quantum electrodynamics 
we thus obtain reliable, well-proven results such as the  Lamb shift 
and the running of the electromagnetic coupling constant due to vacuum polarization. The renormalization process always involves an arbitrary mass parameter $\mu$ and the possibility of rescaling it.  There is also much arbitrariness in the choice of the finite part of the renormalization counterterms. This is also reflected in the predicted running of the coupling constant.  However, when the masses can be neglected  the leading order beta function is uniquely fixed and one obtains 
  $\beta_e \sim  e^3/12\pi^2$ for large $\mu/m$.  In general, when masses are not negligible, the  beta function inherits a dependence on the chosen subtraction scheme. \\
    
Another relevant feature of renormalization is the expected decoupling of higher massive particles, as enforced by the Appelquist-Carazzone theorem \cite{APtheorem}. This means that particles with mass higher than the relevant physical energy scale should not contribute to any computed observable. This ensures that for low energy physics we do not need to know about the related very high energy physics, hence  supporting the effective field theory framework. The minimal subtraction (MS) scheme  in dimensional regularization \cite{Hooft2, Hooft1} is a very efficient method used to evaluate the behavior of the running couplings.  However, MS does not fulfill the decoupling theorem and one needs to resort to a mass-dependent scheme to  capture the low energy behavior of the beta function. \\

 Renormalization theory has also been extended to quantized fields in curved spacetime from  the early 1970s, as reported in  \cite{parker-toms, birrell-davies}. Here the main focus was the renormalization of the energy-momentum tensor and the evaluation of the effective action in a way consistent with general covariance. One of the major tools is the heat-kernel or proper-time expansion of the Feynman propagator \cite{Schwinger, DeWittbook,dewitt75}. As in the case of perturbative computations in Minkowski space, quantized fields in curved space are  also plagued with ultraviolet divergences. The DeWitt-Schwinger  expansion serves to  identify  the  emerging ultraviolet  divergences, some of which are intrinsically tied to the spacetime curvature  and are absent in flat space. In the evaluation of the renormalized effective action the removal of the divergences can also be done using a mass independent scheme, like MS in dimensional regularization \cite{bunch}.  This introduces the usual  $\mu$ parameter and  the associate running of the gravitational coupling constants (see, for instance, \cite{parker-toms, Sola}). As expected,  the obtained runnings  do not fulfill the Appelquist-Carazzone theorem and in consequence make it difficult to arrive at any physical interpretation in the cosmic infrared  regime.   This is specially important in discussing the cosmological constant problem  and the  running of Newton's constant \cite{Sola, Martin, Carroll}. \\
 
 In this work we propose to reevaluate the effective action, and the associated beta functions, by reexpressing  the conventional DeWitt-Schwinger adiabatic expansion with the  introduction of   a novel $\mu$ scale parameter in the definition of the adiabatic subtraction terms. The $\mu$ parameter is  introduced in such a way that a  natural decoupling emerges in the running couplings. 
 We also show how the vacuum metamorphosis model  \cite{parker-raval,parker-ravalA}, one of the most appealing models to account for dark energy \cite{parker-vanzella, caldwell-komp-parker-vanzella} and to soften the measured $H_0$ tension \cite{divalentino}, emerges when the $\mu$ parameter is interpreted in terms of the Ricci scalar. \\

 
 To make the paper self-contained we first  introduce  the DeWitt-Schwinger (proper-time) expansion and  briefly summarize the derivation of the  well-known running for the couplings in  dimensional regularization with the minimal prescription. 
 To better explain the main ideas we consider a quantized complex scalar field coupled to  external  gravitational and electromagnetic fields. The introduction of the external electromagnetic field is somewhat tangential to the main topic of the paper. However, we introduce it in the discussion for pedagogical purposes, since the running of the effective electric charge is a well-established  theoretical and experimental result. 
 This permits one to compare the one-loop electromagnetic behavior with  analogous results in gravity.  
We use units for which $c=1=\hbar$. Our sign conventions for the signature of the metric and the curvature tensor follow Refs. \cite{parker-toms, birrell-davies}.

\section{Effective Action, DeWitt-Schwinger expansion, and minimal subtraction.}
 We start from the classical Einstein-Maxwell theory
\be
S=\int d^4x \sqrt{-g}\left(-\Lambda+\frac{R}{16\pi G}-\frac{1}{4q^2}F_{\mu\nu}F^{\mu\nu}\right)+S_{\rm M} \label{sc}\ 
\ee
coupled to a quantized charged scalar field described by the action
\be
S_{\rm M}=\int d^4x \sqrt{-g}\left((D_{\mu}\phi)^{\dagger}D^{\mu}\phi+m^2|\phi|^2+\xi R |\phi|^2\right) 
\ , \ee
with $D_{\mu}=\nabla_{\mu}+i A_{\mu}$. The most relevant physical objects  are the renormalized energy-momentum tensor $\langle T_{\mu\nu}\rangle$ and the one-loop effective action $S_{\rm eff}$ for the matter field, related by 
$
\frac{2}{\sqrt{-g}}\frac{\delta S_{\rm eff}}{\delta g^{\mu\nu}}=\langle T_{\mu\nu}\rangle \label{Teff} \ .
$
 The effective action can be formally expressed in terms of the  Feynman propagator $
 S_{\rm eff}= -i \operatorname{Tr} \log (-G_F). $ 
The propagator satisfies the Klein-Gordon type equation
\be (\Box_x  + m^2 + \xi R) G_{\rm F}(x, x') = -|g(x)|^{-1/2} \delta (x-x') \ .\ee
In general, the above formal expression for the effective action is divergent. To explicitly identify the ultraviolet divergences, one can express the Feynman propagator as an integral in the proper time $s$ 
\be \label{GFs} G_{\rm F}(x, x') = -i \int_0^\infty ds \ e^{ -im^2 s} \langle x, s |  x', 0\rangle \ , \ee 
where $m^2$ is understood to have an infinitesimal negative imaginary part ($m^2\equiv m^2 -i\epsilon$). 
The kernel $\langle x, s |  x', 0\rangle$ 
can be expanded in powers of the proper time as follows 
 {\be \label{hks}\langle x, s |  x', 0\rangle =  i\frac{\Delta^{1/2} (x, x')}{(4\pi)^2(is)^2}  \exp {\frac{\sigma(x, x')}{2is}} \sum_{j=0}^\infty a_j (x, x')(is)^j \ee 
[$\Delta(x, x')$ is the Van Vleck-Morette determinant and $\sigma(x, x')$ is  the proper distance along  the geodesic from $x'$ to $x$]. Therefore, the effective Lagrangian, defined as $S_{\rm eff}=\int d^4x \sqrt{-g}L_{\rm eff}$, has the following  asymptotic expansion 
\be
L_{\rm eff}=\frac{2i}{2(4\pi)^{2}}\sum^{\infty}_{j=0}a_j(x)\int^{\infty}_0e^{-is m^2}(is)^{j-3}ds \label{Leff}\ .
\ee
The first coefficients $a_n(x, x')$ are given, in the coincidence limit $x \to x'$, by  \cite{parker-toms, birrell-davies}
\bea
a_0(x)=&&1 \ ,  \ \ \ \  \ \ a_1(x)=-\bar{\xi}R \nonumber\\
a_2(x)=&&\frac{1}{180}R_{\alpha\beta\gamma\delta}R^{\alpha\beta\gamma\delta}-\frac{1}{180}R^{\alpha\beta}R_{\alpha\beta}\nonumber \\&&-\frac{1}{6}\left(\frac15-\xi\right)\Box R+\frac{1}{2}\bar{\xi}^2 R^2-\frac{1}{12}F^{\mu\nu}F_{\mu\nu}  \ , \label{coef} 
 \eea
where $\bar{\xi}=\xi-\frac16$. We remark that all dependence on the mass is factored out in the exponential in (\ref{GFs}) (or, equivalently, in (\ref{Leff})). 
 Furthermore, all DeWitt-Schwinger coefficients $a_n$ are polynomial functions of the basic objects: curvature tensors $R_{\alpha\beta\gamma\delta}$, $F_{\mu\nu}$ (and their covariant derivatives), and the metric tensor $g^{\mu\nu}$.  The removal of divergences is usually done via dimensional regularization and minimal subtraction. \\

In $n$ spacetime dimensions the corresponding expression \eqref{Leff}  can be expanded as
\be
L_{eff}\approx\frac{2i}{2(4\pi)^{n/2}}\left(\frac{m}{\mu}\right)^{n-4}\sum^{\infty}_{j=0}a_j(x)m^{4-3j}\Gamma(j-\frac{n}{2}) \ 
\ , \ \ \ \ee
where one has introduced an arbitrary mass scale $\mu$  to maintain the initial units of $L_{\rm eff}$ as $\text{(length)}^4$. $\mu$ is an arbitrary scale, totally independent of $m$. As $n\to4$, the first three terms diverge with simple poles in $1/(n-4)$.
Subtracting the terms with poles one obtains an asymptotic expression for the renormalized effective Lagrangian. This also requires that the original classical Lagrangian be modified, up to total derivatives, by the addition of higher derivative terms of the form $\alpha_1C^2 + \alpha_2R^2$, where $\alpha_1$ and $\alpha_2$ are dimensionless coupling constants. Here $C^2\equiv R_{\mu\nu\alpha\beta}R^{\mu\nu\alpha\beta}-2R_{\mu\nu}R^{\mu\nu}+\frac13R^2$ is the square of the Weyl tensor. Demanding that the total effective Lagrangian, including the classical part,  be $\mu$ independent leads to the following beta functions (see for instance \cite{Sola}) 
\bea &&\beta_{\Lambda}^{MS}=\frac{m^4}{16\pi^2} \ \ \ \ \
\beta_{\kappa}^{MS}=-\frac{m^2\bar{\xi}}{4\pi^2}\ \ \ \ \ \beta_q^{MS}=\frac{q^3}{48\pi^2} \ \ \ \ \ \nonumber \\
&&
 \beta_{\alpha_1}^{MS}=-\frac{1}{960\pi^2}~~~~\beta_{\alpha_2}^{MS}=-\frac{1}{16\pi^2}\bar{\xi}^2 \ \ \ , \label{betadr}
\eea
where $\kappa^{-1}=8\pi G$.  
 The unsatisfactory point of the above results  is the absence of decoupling  for heavy massive fields. \\
\section{Adiabatic DeWitt-Schwinger subtractions. Massless case. }

The DeWitt-Schwinger expansion can also be regarded as an adiabatic  expansion in number of derivatives of the metric and the external  fields.  This is even more explicit in its   counterpart  expansion in  local-momentum space \cite{bunch-parker}.  The high frequency behavior of the Feynman propagator is captured by the DeWitt-Schwinger expansion, irrespective of the background dynamics.
Therefore, the  renormalization of the effective action can also be performed simply by subtracting off all (DeWitt-Schwinger) terms up and including the fourth adiabatic order \cite{parker-toms}
\be
L_{div}=\frac{2i}{2(4\pi)^{2}}\sum^{2}_{j=0}a_j(x)\int^{\infty}_0e^{-is m^2}(is)^{j-3}ds \label{Ldiv}\ .
\ee
However, as stressed in \cite{waldbook}, the DeWitt-Schwinger subtractions are in general  ill defined for $m=0$, due to an infrared divergence in the integration of the heat kernel. 
More precisely, the DeWitt-Schwinger representation of the Feynman propagator can be regarded as a special case of the Hadamard expansion, corresponding to a particular choice of the undetermined biscalar coefficient $\omega_0$ in the Hadamard representation \cite{waldbook}. The DeWitt-Schwinger expansion corresponds to the choice $\omega_0= \omega_0^M + \alpha a_1+ \frac{a_2}{m^2} + \frac{a_3}{m^4} \cdots $. $\omega_0^M$ is the constant value in Minkowski space (see for instance \cite{Had2}). In some special situations, as in the evaluation of  trace anomalies, one can bypass this potential problem by taking the massless limit  at the end of the calculation \cite{birrell-davies}. The result turns out to be finite. 

Here we take a different route. When $m=0$ one can alternatively bypass this infrared issue by introducing a mass scale parameter $\mu$. It can also serve  to define the necessary (but arbitrary) renormalization point. We note that this is somewhat similar to the introduction of the arbitrary length scale $\lambda \sim 1/\mu$ in the  logarithm term $V(x, x') \log \frac{\sigma(x, x')}{\lambda^2} $  of the Hadamard expansion \cite{Had1}. For massive fields, the natural length scale is $\lambda \sim m^{-1}$. However, for massless fields one is forced to introduce the arbitrary scale $\lambda$.  \\

In the DeWitt-Schwinger expansion one can naturally replace the mass parameter $m^2$ in (\ref{Ldiv}) by an arbitrary $\mu^2$ parameter and redefine the DeWitt coefficients $a_i \to \bar a_i$ to keep consistency within each  adiabatic order. The new proposed  $L_{div}(\mu)$ reads
\be
L_{div}(\mu)=\frac{2i}{2(4\pi)^{2}}\sum^{2}_{j=0}\bar{a}_j(x)\int^{\infty}_0e^{-is \mu^2}(is)^{j-3}ds \ , \label{Ldifmu}
\ee
 where the first coefficients $\bar a_i$ of the expansion are 
\bea
&&\bar{a}_0(x)=1 \ , \ \ \ \ \  \ \bar{a}_1(x)=a_1(x)+\mu^2\nonumber\\
&&\bar{a}_2(x)=a_2(x)+\bar{\xi}R \mu^2+\frac12\mu^4 \label{coef1} \ .
\eea

Now we can separate from expression \eqref{Ldifmu} a $\mu$-independent divergent term and a finite $\mu$-dependent term by computing the finite expression
\bea
L_{div}(\mu)-L_{div}(\mu_0)=\delta_{\Lambda}+\delta_{G}R+\delta_{\sigma}a_2
\ , \label{mum}
\eea
where $\delta_{\Lambda}=\frac{-1}{(8\pi)^2}(\mu^4-\mu_0^4); \delta_G=\frac{1}{16\pi^2}\bar{\xi}(\mu^2-\mu_0^2) and \delta_{\sigma}=\frac{-1}{16\pi^2}\log(\mu^2/\mu_0^2)$.
A consequence of the  introduction of the arbitrary scale $\mu$  is the natural emergence of the renormalization group flow \cite{Coleman}.
The beta functions are obtained by requiring $\mu$ independence of the  effective  Lagrangian
\bea L_{eff} &=& -\Lambda(\mu) +\frac{1}{2} \kappa(\mu) R -\frac{1}{4q^2(\mu) }F_{\mu\nu}F^{\mu\nu} \nonumber \\ &+& \alpha_1(\mu)C^2 
+ \alpha_2(\mu) R^2 + \alpha_3(\mu) E + \alpha_4 (\mu) \Box R\nonumber \\
&-& (\delta_{\Lambda}(\mu)+\delta_{G}(\mu)R+\delta_{\sigma}(\mu)a_2) + \cdots \ . \label{effectiveL}\eea
$E=R_{\mu\nu\alpha\beta}R^{\mu\nu\alpha\beta}-4R_{\mu\nu}R^{\mu\nu}+R^2$ is the integrand of the Gauss-Bonet topological invariant. Note that the omitted terms in the third line of (\ref{effectiveL})   are independent of $\mu$. 
The results for the beta functions are 

 \bea
&&\beta_{\Lambda}^{}=\frac{\mu^4}{16\pi^2} \ \ \ \ \ \ \beta_{\kappa}^{}=\frac{\bar{\xi}\mu^2}{4\pi^2} \nonumber \\
&&
\beta_q^{}=\frac{q^3}{48\pi^2} \ \ \ \ \beta_1^{}=\frac{-1}{960\pi^2} \ \ \ \ \beta_2^{}=\frac{-\bar{\xi}^2}{16\pi^2} \nonumber \\ 
&&\beta_3^{}= \frac{1}{2880\pi^2}  \ \ \ \ \ \ \beta_4^{}=\frac{\frac15-\xi}{48\pi^2}\  .    \label{beta}
\eea

We have included for completeness all coupling constants. 
This agrees with the results obtained in \cite{FN} for Friedmann-Lemaitre-Robertson-Walker spacetimes using a similar generalization of the  usual adiabatic regularization method \cite{adiabatic}, via the introduction of an analogous off-shell scale $\mu$.  (For a recent use of this generalization see \cite{moreno-sola}). We also have exact agreement for  the  dimensionless coupling constants  obtained from MS, as displayed in  (\ref{betadr}). Hadamard renormalization also leads to a similar result for the running of the electric coupling constant \cite{Had1, BNP}. \\

 \section {Massive case, decoupling and running gravitational constants}

Now we want to generalize the previous analysis to massive fields. 
Therefore, instead of (\ref{Ldifmu}) and  (\ref{Ldiv}) we should write
 \be
L_{div}(\mu)=\frac{2i}{2(4\pi)^{2}}\sum^{2}_{j=0}\bar{a}_j(x)\int^{\infty}_0e^{-is f(\mu^2, m^2)}(is)^{j-3}ds \ . \label{Ldifmu2}
\ee
 The simplest choice for the   function $f(\mu^2, m^2)$ is 
$ f(\mu^2, m^2) = m^2 + \mu^2$.
This choice is univocally singularized if we demand that the  mass $m^2$ is factored out in the form of an exponential $e^{-ism^2}$, as in the conventional DeWitt-Schwinger expansion (\ref{Ldiv}). Furthermore,  for $m=0$ we have to recover (\ref{Ldifmu}). Hence

\be
L_{div}(\mu)=\frac{2i}{2(4\pi)^{2}}\sum^{2}_{j=0}\bar{a}_j(x)\int^{\infty}_0e^{-is(m^2 + \mu^2)}(is)^{j-3}ds \ , \label{Ldifmu2}
\ee
$\bar{a}_0(x)=1$,$ \bar{a}_1(x)=a_1(x)+\mu^2$, and $\bar{a}_2(x)=a_2(x)+\bar{\xi}R\mu^2+\frac12\mu^4$ are kept mass independent. Note that any other expression for $f$ implies dependence on the mass of the redefined DeWitt coefficients $\bar a_i$. Even more, any other choice for $f$ implies a nonpolynomial dependence of the coefficients $\bar a_i$ on $m^2$. \\

 The corresponding beta function for the electric charge obtained from (\ref{Ldifmu2})  is
 \be
\beta_q=\frac{q^3}{48\pi^2}\frac{\mu^2}{m^2+\mu^2} \label{betaq2}\ ,
\ee
while the result for the dimensionless gravitational constants are similarly 
\bea \label{betaalpha}
\beta_1=-\frac{1}{960\pi^2}\frac{\mu^2}{m^2+\mu^2}~~~~\beta_2=-\frac{\bar{\xi}^2}{16\pi^2}\frac{\mu^2}{m^2+\mu^2}\nonumber\\ \beta_3=\frac{1}{2880\pi^2}\frac{\mu^2}{m^2+\mu^2}   ~~~~\beta_4=\frac{\frac15-\xi}{48\pi^2}\frac{\mu^2}{m^2+\mu^2} \ .
\eea
The difference between \eqref{betaq2}-\eqref{betaalpha} and \eqref{beta} is that the former approaches the latter in the limit $\mu \gg m$ while it approaches to zero quadratically in the limit $\mu \ll m$. This is equivalent to the decoupling of very massive charged particles in scalar electrodynamics.  \\


Concerning the dimensionfull gravitational constants, the decoupling is also absent in dimensional regularization. 
This makes it not trivial to assign some physical meaning to the $\mu$ parameter. However, within the proposed DeWitt-Schwinger framework  and from (\ref{Ldifmu2}) we get the following beta functions:
\be
  \beta_{\Lambda}=\frac{1}{16\pi^2}\frac{\mu^6}{m^2+\mu^2}~~~,\beta_{\kappa}=\frac{\bar{\xi}}{4\pi^2}\frac{\mu^4}{m^2+\mu^2} \ .
\ee
For large values of the scale $\mu \gg m$ the mass can be ignored, while heavy particles $m \gg \mu$ decouple and the beta functions tend to zero. 
Note that the decoupling of the dimensionfull gravitational constants, in contrast with the dimensionless ones,  is a highly nontrivial issue \cite{gorbar, markkanen, babic, franchino19, franchino-review}. \\

The running of the cosmological and Newton's gravitational constants are given by ($\Lambda= \Lambda_c/8\pi G$, where $\Lambda_c$ is the traditional cosmological constant)
\bea \label{rlambda}
\Lambda(\mu)&=&\Lambda_0+\frac{1}{64\pi^2}((\mu^4-\mu_0^4)-2m^2(\mu^2-\mu_0^2) \nonumber \\
&+& 2m^4\log{\left(\frac{m^2+\mu^2}{m^2+\mu_0^2}\right)}) \label{ccmu}  \\
G(\mu)&=&\frac{G_0}{1+\frac{\bar{\xi} G_0}{\pi}\left(\mu^2-\mu_0^2-m^2\log{\left(\frac{m^2+ \mu^2}{m^2+\mu_0^2}\right)}\right)} \ , \label{rG}
\eea
while the running for the dimensionless gravitational constants are
\bea \label{rlambda}
\alpha_i(\mu)&=&\alpha_{i0}+\frac{\sigma_i}{4\pi^2}\log\left( \frac{m^2 + \mu^2}{m^2 + \mu_0^2} \right) \ .
\eea
where $\sigma_1= -\frac{1}{430}$, $\sigma_2=-\frac{\bar \xi^2}{8}$, $\sigma_3=\frac{1}{1440}$, and $\sigma_4=\frac{(1/5-\xi)}{24}$. \\
\section{Relation with other approaches}
 It is interesting to briefly consider the massless limit for the predicted running for the Newton constant, as given by (\ref{rG}): $
G(\mu)=G_0(1+ \frac{\bar \xi}{\pi} \ G_0(\mu^2-\mu_0^2))^{-1}$.
This expression has the same form as the one obtained within a very different approach. The asymptotic safety framework of quantum gravity predicts a similar behavior for the running of Newton's constant \cite{review}
(see also \cite{polyakov}). \\

Even though the above renormalization  prescription  does not give us a uniquely physical interpretation for $\mu$,  it supports the idea that indeed it can be linked to some physical scale, such as the conventional momentum scaling  $p_i \to s p_i$  in flat space particle scattering associated with the scaling $\mu \to s \mu_0$. In curved spacetime the scaling of $\mu$ should be linked, by dimensional reasons, to the scaling of the metric $g_{\mu\nu} \to s^{-2} g_{\mu\nu}$, and hence to the scaling of the curvature $R \to s^2 R$ \cite{Nelson}. Therefore, while the dependence on $\mu$ of the renormalized electric charge  has the same form as the dependence of the measured charge on the square of the momentum transfer in electron scattering, the dependence of the renormalized $\Lambda$ or $\kappa$ on $\mu$ is expected to be traded to the curvature dependence of the observable  gravitational constants. One possible way of choosing a natural mass/length scale in a cosmological setting is to make $\mu$ proportional to the Hubble parameter $H$, or $\mu^2$ to be proportional to the Ricci scalar  $R$. Here we are more interested in the infrared behavior of the runnings, and hence in the low curvature regime. The runnings obtained above are somewhat similar to the generic form of the running proposed in the  running vacuum models \cite{Sola-Shapiro} (see also \cite{Sola, moreno-sola} and \cite{Sola2} for a connection with cosmological observations and smoothing of data tensions). The connection between  $\mu^2$ and $R$ has also been previously suggested in \cite{markkanen}. \\


  Let us analyze with more details the consequences of the assumption $\mu^2 \propto R$.  For computational purposes it is convenient to choose $\mu^2=\bar{\xi}R$. We also select the reference point $\mu_0=0$ and assume that
\be \label{iconditions} \Lambda_0=0, \ \ \  \alpha_{i 0}=0 \ , \ee
and keep $\kappa_0= (1/8 \pi G_0)$, where $G_0$ is the measured Newton's constant. These renormalization conditions can be understood as  our definition of the physical gravitational constants in the very infrared limit point. The effective Lagrangian is well approximated,  in the adiabatic limit of our late-time expanding Universe, by (here we are considering a single  real scalar field)
\bea L_{eff} &= & -\Lambda(\mu) +\frac{1}{2} \kappa(\mu) R + \alpha_1(\mu)C^2 + \alpha_2(\mu) R^2 +\nonumber \\ &+&  \alpha_3(\mu) E + \alpha_4 (\mu) \Box R \ . \eea
Taking into account the running derived in (\ref{ccmu})-(\ref{rlambda}) for all  gravitational coupling constants and the conditions (\ref{iconditions}), the above effective action can be rewritten in the form 
\bea \label{PRmodel}
 L_{eff} &&= \frac12 \kappa_0 R+\frac{1}{64\pi^2}\Big\{m^2 \bar{\xi} R+\frac32 \bar{\xi}^2 R^2\nonumber \\&&-\left(m^4+ 2m^2  \bar{\xi} R+2a_2\right)\log{\left(\frac{m^2 + \bar \xi R}{m^2}\right)}\Big\} \ .
\
\eea 
Remarkably, this coincides with the  action proposed  by Parker and Raval in \cite{parker-raval,parker-ravalA}, and known as the vacuum metamorphosis model \cite{parker-vanzella, caldwell-komp-parker-vanzella}, on the basis of the $R$-summed form of the Feynman propagator \cite{Parker-Toms85, Jack-Parker, FNP}.
Here only the measured Newton's constant $G_0$ appears in the action. 
The semiclassical  dynamics of (\ref{PRmodel}) provides negative pressure  to suddenly accelerate the Universe at a rate compatible with observations (it softens also the $H_0$ tension \cite{divalentino}) for an ultralow mass scalar field, of the same order as the current expansion rate of the Universe ($m\sim H_0$), in accordance with the underlying decoupling mechanism displayed above. This provides further evidence for the connection between the parameter $\mu$ in our proposed physical renormalization scheme with the physical scale $R$.

\section{Conclusions and final comments}   

We have generalized the DeWitt-Schwinger  renormalization subtractions  to include an arbitrary renormalization mass scale $\mu$, and in such a way to ensure the decoupling of heavy masses. This is a somewhat improvement with respect to the more common treatment of the DeWitt-Schwinger expansion via dimensional regularization and minimal subtraction. We have also analyzed the new predicted running for the gravitational couplings. \\

As a by-product of our proposal, and because of the natural decoupling,  the obtained runnings could be of interest for the issue of the 
  cosmological constant problem.   
To see this in the conventional   way  let us assume that $\Lambda_0=0$. Following the standard approach, i.e., dimensional regularization and MS, any massive particle will contribute as $
 \Lambda^{\bf MS}(\mu) \sim m^4 \log{\left(\frac{\mu^2}{m^2}\right)} 
 $ (see, for instance, \cite{Martin, Sola}) and taking the characteristic scale of the Standard Model  gives the well-known extremely high contribution $\Lambda \sim 10^{46}\text{eV}^4$. This is in conflict with  the observed current energy density $\Lambda_{\bf obs}\sim 10^{-11}\text{eV}^4$ (see \cite{Sola} for a detailed discussion). However, if we now use \eqref{ccmu} we obtain an extremely low value. 
More generally, in the limit of large masses  $m\gg\mu \sim H_0$ (all the standard model particles) the term $m^4$ decouples and we get
$
 \Lambda^{DS}(\mu) \sim \frac{\mu^6}{m^2}+\mathcal{O}(\frac{1}{m^4})  
 $. 
This heuristic discussion suggests that the origin of the accelerated expansion could be more  naturally found in ultralow masses.  
This requires the identification of $\mu^2$ as a time-dependent scale proportional to the Ricci scalar,  as also reinforced in the more quantitative arguments displayed in this work. Further work is required to make more definite statements.\\



{\it Acknowledgments.--} We thank P. Beltran-Palau,      S. A. Franchino-Vinas, S. Nadal and S. Pla for useful comments. This work has been supported by the Spanish MINECO research grants No. FIS2017- 84440-C2-1-P and No. FIS2017-91161-EXP.  A. F. is supported by the Severo Ochoa Ph.D. fellowship, Grant No. SEV-2014-0398-16-1, and the European Social Fund.}


\begin{thebibliography}{99}


\bibitem{qftbook1} S. Weinberg, {\it The Quantum Theory of Fields}, Vol. 1,2,  Cambridge University Press, Cambridge, (1995). 

\bibitem{qftbook2} M. E. Peskin and D. V.  Schroeder, {\it An Introduction to Quantum Field Theory}, Addison-Wesley, Reading MA, (1995).

\bibitem{qftbook3} L. Alvarez-Gaume and M.A. Vazquez-Mozo, {\it An Invitation to Quantum Field Theory}, Springer-Verlag, Berlin,  (2012).

\bibitem{APtheorem} T. Appelquist and J. Carazzone, {\it Phys. Rev. D} {\bf 11}, 2856 (1975).



\bibitem{Hooft2} G.~'t Hooft, {\it Nucl. Phys. B} {\bf 61}, 455 (1973).


\bibitem{Hooft1} G.~'t Hooft and M.~J.~G.~Veltman,
  Nucl.\ Phys.\ B {\bf 44} (1972) 189.







  



\bibitem{parker-toms}L.~Parker and D.~J.~Toms, {\it Quantum Field Theory in Curved Spacetime: Quantized Fields
and Gravity}, Cambridge University Press, Cambridge, England (2009).

\bibitem{birrell-davies} N.~D.~Birrell  and P.~C.~W.~Davies, {\it Quantum Fields in Curved Space}, Cambridge University Press, Cambridge, England (1982).






\bibitem{Schwinger} J. Schwinger, {\it Phys. Rev.}{\bf 82}, 664 (1951).

\bibitem{DeWittbook} B. S. DeWitt, {\it Dynamical theory of groups and fields}, Gordon and Breach, New York (1965). 


\bibitem{dewitt75} B. S. DeWitt, {\it Phys. Rep.} {\bf 19}, 295 (1975).

\bibitem{bunch}T. S. Bunch, {\it J. Phys. A} {\bf 12}, 4 (1979).

\bibitem{Sola} J. Sol{\`{a}}, 
{\it J. Phys. Conf. Ser.} {\bf 453}, 012015 (2013). 
\bibitem{Martin} J. Martin, 
    {\it Comptes Rendus Physique} {\bf 13}, 566-665 (2012).
 \bibitem{Carroll} S. M. Carroll, {\it Living Rev. Relativ.} {\bf 4}, 1 (2001).
 
 
 
 
 
 

 
 \bibitem{parker-raval} L. Parker and A. Raval, {\it Phys. Rev. D} {\bf 60}, 063512 (1999). 
\bibitem{parker-ravalA} L. Parker and A. Raval,  {\it Phys. Rev. D} {\bf 62}, 083503 (2000); {\it Phys. Rev. Lett.} {\bf 86}, 749 (2001).


\bibitem{parker-vanzella} L. Parker and D. A. T. Vanzella, {\it Phys. Rev. D} {\bf 69}, 104009 (2004).
\bibitem{caldwell-komp-parker-vanzella} R.R. Caldwell, W. Komp, L. Parker and D.A.T. Wanzella, {\it Phys. Rev. D}{\bf 73}, 023513 (2006).

\bibitem{divalentino}E. Di Valentino, E. V. Linder, A. Melchiorri,
{\it Phys.Rev. D} {\bf 97}, 4 (2018).

 

\bibitem{bunch-parker} T. S. Bunch and L. Parker, {\it Phys. Rev. D} {\bf 20}, 2499 (1979).
\bibitem{waldbook} R. M. Wald, {\it Quantum Field Theory in Curved Spacetime and Black Hole Thermodynamics}, University of Chicago Press, Chicago (1994).

\bibitem{Had2} M. A. Castagnino and D. D. Harari, {\it Ann. Phys. (NY)} {\bf 152}, 85 (1984).

 \bibitem{Had1}V. Balakumar, E. Winstanley, {\it Class. Quant. Grav.} {\bf 37},  065004 (2020).


\bibitem{Coleman} S. Coleman, {\it Aspects of Symmetry}, Cambridge University Press, Cambridge (1985). 

\bibitem{FN}A. Ferreiro and J. Navarro-Salas, {\it Phys. Lett. B} {\bf 792}, 81 (2019).

\bibitem{adiabatic}  L.~Parker and S.~A.~Fulling, {\it Phys.~Rev.~D} {\bf 9}, 341 (1974). P.~R.~Anderson and L.~Parker, {\it Phys.~Rev.~D} {\bf 36}, 2963 (1987).  I. Agullo, J. Navarro-Salas, G. J. Olmo, and L. Parker,  {\it Phys.Rev. D} {\bf 84},  107304 (2011).

 \bibitem{moreno-sola}     C. Moreno-Pulido and J. Sola, {\it Running vacuum in quantum field theory in curved spacetime: renormalizing $\rho_{vac}(H)$ without $\sim m^4$ terms}, arXiv:     2005.03164.
   
\bibitem{BNP} P. Beltran-Palau, J. Navarro-Salas and S. Pla, {\it Phys. Rev. D} {\bf 101}, 105014 (2020).


\bibitem{gorbar} E. V. Gorbar and I.L. Shapiro, {\it JHEP}\ 02, (2003) 021.

\bibitem{markkanen} T. Markkanen, {\it Phys. Rev. D} {\bf 91}, 124011 (2015).
 
 \bibitem{babic} A. Babic, B. Guberina, R. Horvat and H.
Stefancic, {\it  Phys. Rev. D} {\bf 65},  085002 (2002) ; {\it  Phys. Rev. D} {\bf 71}, 124041 (2005).

\bibitem{franchino19} S. A. Franchino-Vinas, T. D.  Netto, I.L. Shapiro, O. Zanusso, {\it Phys. Lett. B} {\bf 790}, 229  (2019).

\bibitem{franchino-review}     S. A. Franchino-Vinas, T. de Paula Netto, O. Zanusso, {\it Universe} {\bf 5},  (2019) 3, 67.


\bibitem{review} M. Niedermaier and M. Reuter, {\it Living Rev. Rel.} {\bf 9}, 5 (2006). M. Reuter,   	
{\it Newton's constant isn't constant}; arXiv: hep-th/0012069.


\bibitem{polyakov} A. Polyakov, in {\it Gravitation and Quantization}, J. Zinn-Justin, B. Julia
(Eds.), North-Holland, (1995).
 \bibitem{Nelson} B. Nelson and P. Panangaden, {\it Phys. Rev. D} {\bf 25}, 1019 (1982). L. Parker and D. Toms, {\it Phys. Rev. D} {\bf 29}, 1584 (1984). S. Hollands and R. M. Wald, {\it Commun. Math. Phys.} {\bf 237}, 123 (2003).

\bibitem{Sola-Shapiro}  I.  L.  Shapiro  and  J.  Sol{\`{a}},  {\it JHEP} 02, (2002) 006; {\it Phys. Lett. B} {\bf 475},   236 (2000); {\it Phys. Lett. B} {\bf 682}, 105 (2009).
\bibitem{Sola2} J. Sol{\`{a}}, A. G{\'{o}}mez-Valent and J. de Cruz P{\'{e}}rez, {\it Astrophys J.}, 836, 43 (2017). J. Sol{\`{a}} , Int. J. Mod. Phys. A, 33, 1844009 (2018).



\bibitem{Parker-Toms85} L.~Parker and D.~J.~Toms, {\it Phys. Rev. D} {\bf 31}, 953 (1985).

\bibitem{Jack-Parker} I. Jack and L. Parker, {\it Phys. Rev. D} {\bf 31}, 2439 (1985).

\bibitem{FNP} A. Ferreiro, J. Navarro-Salas and S. Pla, {\it Phys. Rev. D} {\bf 101}, 105011 (2020).







    
    




























    
 
    
      
      
      


 




























---------------------------------------------

  \end{thebibliography}
\end{document}